\documentclass[12pt]{iopart}

\usepackage{cite}

\usepackage{graphicx}

\usepackage{dcolumn}

\begin{document}

\title{On the uncertainty relations for an electron in a constant magnetic field}
\author{Francisco M. Fern\'{a}ndez}

\address{INIFTA (UNLP, CCT La Plata-CONICET), Divisi\'on Qu\'imica Te\'orica,
Blvd. 113 S/N,  Sucursal 4, Casilla de Correo 16, 1900 La Plata,
Argentina}\ead{fernande@quimica.unlp.edu.ar}

\maketitle

\begin{abstract}
We discuss the uncertainty relation for the azimuthal angle $\phi$
and the $z$-component of the angular momentum $L_z$. To this end
we derive the uncertainty relation for an arbitrary pair of
observables and discuss the conditions for its validity. By means
of a simple parameter-dependent state we illustrate the well known
fact that the standard uncertainty relation for the coordinate and
its conjugate momentum does not apply the pair of observables
$\phi-L_z$. This analysis is motivated by a discussion of the
motion of an electron in a constant magnetic field appeared
recently in this journal (Eur. J. Phys. \textbf{33} (2012) 1147)
where the author assumed the validity of the standard uncertainty
relation for the pair $\phi-L_z$.
\end{abstract}

\section{Introduction}

\label{sec:intro}

In a recent paper Strange\cite{S12} discussed some quantum-mechanical
properties of an electron in a constant magnetic field. Since the system is
axially symmetric along the field direction (chosen to be the $z$ axis) then
the projection of the angular momentum along that axis is a constant of the
motion. The motion of the electron is free along the $z$ axis and bounded on
the plane $x-y$. Restricting the motion of the electron to this plane
Strange discussed the uncertainty relation for the azimuthal angle $\phi $
and the $z$--component of the angular momentum $L_{z}$ that he assumed to be
$\Delta \phi \Delta L_z \ge \hbar$. However, he did not take into account
some of the subtleties of this uncertainty relation that make it quite
different from that for a cartesian coordinate and its conjugate linear
momentum. The $\phi -L_{z}$ uncertainty relation was discussed by several
authors in the past\cite{J64,K65,PT69,K70,P79,C01}. There is even an
interesting series of pedagogical articles on the subject\cite{PT69, K70,
P79, C01}, not without some controversy\cite{PT69, K70}. According to those
papers the uncertainty relation invoked by Strange is incorrect. For this
reason we deem it worthwhile to carry out a more detailed analysis of the
results derived by this author, particularly because the $\phi-L_{z}$
uncertainty relation is suitable for an undergraduate course on quantum
mechanics\cite{PT69, K70, P79, C01}.

In section \ref{sec:uncertainty} we derive the uncertainty
relation for an arbitrary pair of observables following
Chisolm\cite{C01}. In section \ref {sec:example} we first outline
Strange's results based on the incorrect $\phi -L_{z}$ uncertainty
relation and then derive an exact one following
Kraus\cite{K65,K70} and Chisolm\cite{C01}. We also contrast the
exact uncertainty relation with the incorrect one by means of a
state that is somewhat more general than the one chosen by
Strange. Finally, in section \ref{sec:conclusions} we summarize
the main results of this paper and draw conclusions.

\section{The uncertainty relations}

\label{sec:uncertainty}

In order to make this paper sufficiently self-contained and
facilitate the discussion of the uncertainty relation for the
electron in a constant magnetic field\cite{S12} in what follows we
derive the uncertainty relation for an arbitrary pair of
observables. There are different ways of deriving it\cite{P79,C01}
and in what follows we resort to the well known Schwarz
inequality\cite{C01}. To this end consider the usual complex inner
product in quantum mechanics in terms of the bra-ket notation:
$\left\langle f\right| \left. g\right\rangle =\left\langle
g\right| \left. f\right\rangle ^{*}$. The Schwarz inequality
states that
\begin{equation}
\left| \left\langle f\right| \left. g\right\rangle \right| ^{2}\leq
\left\langle f\right| \left. f\right\rangle \left\langle g\right| \left.
g\right\rangle  \label{eq:schwarz}
\end{equation}
for any two vectors $\left| f\right\rangle $ and $\left| g\right\rangle $ in
the state vector space. Chisolm\cite{C01} derived a somewhat more general
uncertainty relation from the obvious expression
\begin{equation}
\left| \left\langle f\right| \left. g\right\rangle \right| ^{2}=\frac{1}{4}%
\left( \left\langle f\right| \left. g\right\rangle +\left\langle g\right|
\left. f\right\rangle \right) ^{2}+\frac{1}{4}\left| \left\langle f\right|
\left. g\right\rangle -\left\langle g\right| \left. f\right\rangle \right|
^{2}
\end{equation}
However, for present purposes it is sufficient to take into account that
\begin{equation}
\left| \left\langle f\right| \left. g\right\rangle \right| \geq \frac{1}{2}%
\left| \left\langle f\right| \left. g\right\rangle -\left\langle g\right|
\left. f\right\rangle \right|
\end{equation}
(that is to say $\left| \left\langle f\right| \left.
g\right\rangle \right| \geq \left| \mathrm{Im}\left\langle
f\right| \left. g\right\rangle \right| $) that leads to
\begin{equation}
\sqrt{\left\langle f\right| \left. f\right\rangle \left\langle g\right|
\left. g\right\rangle }\geq \frac{1}{2}\left| \left\langle f\right| \left.
g\right\rangle -\left\langle g\right| \left. f\right\rangle \right|
\label{eq:schwarz2}
\end{equation}

Let $\left| \psi \right\rangle $ be the state of the system normalized to
unity ($\left\langle \psi \right| \left. \psi \right\rangle =1$) and $\hat{A}
$ and $\hat{B}$ the Hermitean operators for two quantum-mechanical
observables. We define $\left| f\right\rangle =\left( \hat{A}-\left\langle
\hat{A}\right\rangle \right) \left| \psi \right\rangle $ and $\left|
g\right\rangle =\left( \hat{B}-\left\langle \hat{B}\right\rangle \right)
\left| \psi \right\rangle $, where $\left\langle \hat{Q}\right\rangle
=\left\langle \psi \right| \hat{Q}\left| \psi \right\rangle $, so that
\begin{eqnarray}
\left\langle f\right| \left. f\right\rangle &=&\left\langle \hat{A}%
^{2}\right\rangle -\left\langle \hat{A}\right\rangle ^{2}=\left( \Delta
A\right) ^{2}  \nonumber \\
\left\langle g\right| \left. g\right\rangle &=&\left\langle \hat{B}%
^{2}\right\rangle -\left\langle \hat{B}\right\rangle ^{2}=\left( \Delta
B\right) ^{2}  \label{eq:DeltaA_DeltaB}
\end{eqnarray}
Since $\left\langle f\right| \left. g\right\rangle =\left\langle \hat{A}\psi
\right| \left. \hat{B}\psi \right\rangle -\left\langle \hat{A}\right\rangle
\left\langle \hat{B}\right\rangle $ then it follows from equation~(\ref
{eq:schwarz2}) that
\begin{equation}
\Delta A\Delta B\geq \frac{1}{2}\left| \left\langle \hat{A}\psi \right|
\left. \hat{B}\psi \right\rangle -\left\langle \hat{B}\psi \right| \left.
\hat{A}\psi \right\rangle \right|  \label{eq:uncertaiinty_gen}
\end{equation}
If $\hat{B}\left| \psi \right\rangle $ belongs to the domain of $\hat{A}$
and $\hat{A}\left| \psi \right\rangle $ to the domain of $\hat{B}$ then we
can write
\begin{eqnarray}
\left\langle \hat{A}\psi \right| \left. \hat{B}\psi \right\rangle
&=&\left\langle \psi \right| \hat{A}\hat{B}\left| \psi \right\rangle
\nonumber \\
\left\langle \hat{B}\psi \right| \left. \hat{A}\psi \right\rangle
&=&\left\langle \psi \right| \hat{B}\hat{A}\left| \psi \right\rangle
\label{eq:turn_over}
\end{eqnarray}
and thus obtain the standard uncertainty relation
\begin{equation}
\Delta A\Delta B\geq \frac{1}{2}\left| \left\langle \left[ \hat{A},\hat{B}%
\right] \right\rangle \right|  \label{eq:uncertainty}
\end{equation}
where $\left[ \hat{A},\hat{B}\right] =\hat{A}\hat{B}-\hat{B}\hat{A}$ is the
well known commutator. The interested reader will find a more detailed
discussion of the domains and ranges of operators in the literature already
cited\cite{J64,K65,PT69,K70,P79,C01}.

Before applying the results of this section to a particular model
in the next one it is worth stressing the fact that equation
(\ref{eq:uncertainty}) is valid provided that the root-mean-square
deviations $\Delta A$ and $\Delta B$ are calculated according to
equation~(\ref{eq:DeltaA_DeltaB}) and that equations
(\ref{eq:turn_over}) hold for the chosen state $\left| \psi
\right\rangle $. If the chosen state and operators do not satisfy
the latter conditions we can still use the more general inequality
(\ref {eq:uncertaiinty_gen}).

\section{Uncertainty relation for the azimuthal angle and angular momentum}

\label{sec:example}

Strange\cite{S12} described the motion of the electron in the
$x-y$ plane in polar coordinates $x=r\cos \phi $, $y=r\sin \phi $,
where $0\leq r=\sqrt{x^{2}+y^{2}}<\infty $ and $0\leq \phi <2\pi
$. For simplicity we omit the variable $r$ that is not relevant to
the discussion of the uncertainty relation for $\hat{\phi}$ and
$\hat{L}_{z}$ that commutes with the Hamiltonian operator of the
system. In the coordinate representation we define these operators
as follows:
\begin{eqnarray}
\hat{\phi}\psi (\phi ) &=&\phi \psi (\phi )  \nonumber \\
\hat{L}_{z}\psi (\phi ) &=&-i\hbar \frac{\partial }{\partial \phi }\psi
(\phi )  \label{eq:operators}
\end{eqnarray}
where $\psi (\phi )=\left\langle \phi \right| \left. \psi
\right\rangle $. Although it has been argued that this definition
of the quantum-mechanical operator for the azimuthal angle may not
be correct\cite{J64,PT69,K70} we keep it here because it is
relevant to the discussion of the results obtained by
Strange\cite{S12}. Besides, Chisolm\cite{C01} already chose this
definition of $\hat{\phi}$ in his discussion of the uncertainty
relations. We assume the state vectors to be periodic functions of
period $2\pi $ ($f(\phi +2\pi )=f(\phi )$) and choose the standard
inner product
\begin{equation}
\left\langle f\right| \left. g\right\rangle =\int_{0}^{2\pi }f(\phi
)^{*}g(\phi )\,d\phi  \label{eq:inner_prod_phi}
\end{equation}

Strange\cite{S12} stated that ``The azimuthal angle-angular
momentum uncertainty relation is $\Delta \phi \Delta L\geq \hbar
$''. The origin of this uncertainty relation is unclear as it
differs from the standard one $\Delta x\Delta p\geq \hbar /2$ for
the coordinate $x$ and its conjugate momentum $p$. In order to
verify this uncertainty relation he later chose ``an equally
weighted sum of the $m=0$ and $m=1$ state''. Since he did not
write the state explicitly we suppose that it was of the form
\begin{equation}
\psi _{S}(\phi )=\frac{1}{2\sqrt{\pi }}\left( 1+e^{i\phi }\right)
\label{eq:psi_Strange}
\end{equation}
from which we obtain $\left\langle \hat{L}_{z}\right\rangle =\hbar
/2$, $\left\langle \hat{L}_{z}^{2}\right\rangle =\hbar ^{2}/2$ and
$\Delta L_{z}=\hbar /2$ in agreement with his results. Arguing
that ``the uncertainty in angle arises directly from the fact that
the origin of the angular coordinate is arbitrary'' he chose
$\left( \Delta \phi \right) _{S}=\pi $ and obtained $\left( \Delta
\phi \right) _{S}\Delta L_{z}=\pi \hbar /2$. However, in section
\ref{sec:uncertainty} we showed that the uncertainty relation
(\ref{eq:uncertainty}) is valid if the root-mean-square deviations
are calculated as in equation (\ref{eq:DeltaA_DeltaB}). In the
present case the inequality holds for $\Delta \phi =\sqrt{2+\pi
^{2}/3}$ and, therefore, also for $\left( \Delta \phi \right)
_{S}>\Delta \phi $.

The results just discussed are valid for the particular state (\ref
{eq:psi_Strange}). It is convenient to derive the $\phi -L_{z}$ uncertainty
relation for an arbitrary wave function $\psi (\phi )$ of period $2\pi $. If
we integrate $\left\langle \hat{L}_{z}\psi \right| \left. \phi \psi
\right\rangle $ \thinspace by parts we obtain\cite{C01}
\begin{equation}
\left\langle \hat{L}_{z}\psi \right| \left. \phi \psi \right\rangle
=\left\langle \psi \right| \left. \hat{L}_{z}\phi \psi \right\rangle +i\hbar
2\pi \left| \psi (2\pi )\right| ^{2}
\end{equation}
and equation (\ref{eq:uncertaiinty_gen}) leads to the exact inequality
\begin{equation}
\Delta \phi \Delta L_{z}\geq \frac{\hbar }{2}\left| 2\pi \left| \psi (2\pi
)\right| ^{2}-1\right|  \label{eq:Delta_phi_Delta_Lz}
\end{equation}
already derive earlier by other authors\cite{K65,K70,C01}. The
reason why $\left\langle \hat{L}_{z}\psi \right| \left. \phi \psi
\right\rangle \neq \left\langle \psi \right| \left.
\hat{L}_{z}\phi \psi \right\rangle $ is that $\phi \psi (\phi )$,
unlike $\psi (\phi )$, is not a periodic function of period $2\pi
$ and, consequently, does not belong to the domain of
$\hat{L}_{z}$ (a more detailed discussion of this issue is
available in the articles already
cited\cite{J64,K65,PT69,K70,P79,C01}). However, note that when
$\left| \psi \right\rangle =\left| \psi _{S}\right\rangle $ the
right-hand-side of equation (\ref{eq:Delta_phi_Delta_Lz}) is
exactly $\hbar /2$ because $\left| \psi _{S}(2\pi )\right|
^{2}=1/\pi $. In other words, the `standard' uncertainty relation
$\Delta \phi \Delta L_{z}\geq \hbar /2$ is valid for the
particular wave function $\psi _{S}(\phi )$ chosen by Strange as
an illustrative example.

Since the right-hand side of equation
(\ref{eq:Delta_phi_Delta_Lz}) may be smaller that $\hbar /2$ the
standard uncertainty relation $\Delta \phi \Delta L_{z}\geq \hbar
/2$ is not guaranteed. We think that it is a worthy pedagogical
experiment to test its validity on other state functions. For
example, we can try a more general linear combination of the same
two states with $m=0$ and $m=1$:
\begin{equation}
\psi (a,\phi )=\frac{1}{\sqrt{2\pi }}\left( a+\sqrt{1-a^{2}}e^{i\phi }\right)
\label{eq:Psi(a,phi)}
\end{equation}
where $-1\leq a\leq 1$, which reduces to $\psi _{S}(\phi )$ when
$a=1/\sqrt{2}$. With this simple function we easily obtain
\begin{eqnarray}
R(a) &=&\frac{\hbar }{2}\left| 2\pi \left| \psi (a,2\pi )\right|
^{2}-1\right| =\hbar |a|\sqrt{1-a^{2}}  \nonumber \\
\Delta L_{z} &=&\hbar |a|\sqrt{1-a^{2}}=R(a)  \nonumber \\
\Delta \phi &=&\left( 4a\sqrt{1-a^{2}}+\frac{\pi ^{2}}{3}\right) ^{1/2}
\label{eq:R(a)_DLz_Dphi}
\end{eqnarray}
Note that $\Delta L_{z}=0$ when $a=0$ or $a=1$ because $\psi
(0,\phi )$ and $\psi (1,\phi )$ are eigenfunctions of
$\hat{L}_{z}$, and that in both cases $\Delta \phi =\pi
/\sqrt{3}$. Besides, it follows from $\Delta \phi \Delta L_{z}\geq
R(a)$ that $\Delta \phi \geq 1$ for all $-1\leq a\leq 1$.

\begin{figure}[H]
\begin{center}
\bigskip\bigskip\bigskip \includegraphics[width=9cm]{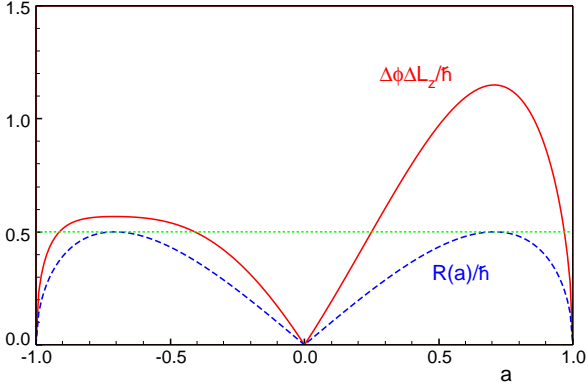}
\end{center}
\caption{$\Delta \phi \Delta L_z/\hbar$ and $R(a)/\hbar$ vs. $a$}
\label{Fig:DphiDLz}
\end{figure}

Fig.~\ref{Fig:DphiDLz} shows that $\Delta \phi \Delta L_{z}=\hbar
/2$ at four points: $a_{1}\approx -0.91$, $a_{2}\approx -0.41$,
$a_{3}\approx 0.25$ and $a_{4}\approx 0.97$. The standard
uncertainty relation $\Delta \phi \Delta L_{z}\geq \hbar /2$ holds
only for $a_{1}\leq a\leq a_{2}$ and $a_{3}\leq a\leq a_{4}$,
while, on the other hand, the exact one $\Delta \phi \Delta
L_{z}\geq R(a)$ is valid for all $a$. In addition to it,
$R(a)=\hbar /2$ only for $a=\pm 1/\sqrt{2}$, that is to say, for
an equally weighted sum of the states with $m=0$ and $m=1$.

\begin{figure}[H]
\begin{center}
\bigskip\bigskip\bigskip \includegraphics[width=9cm]{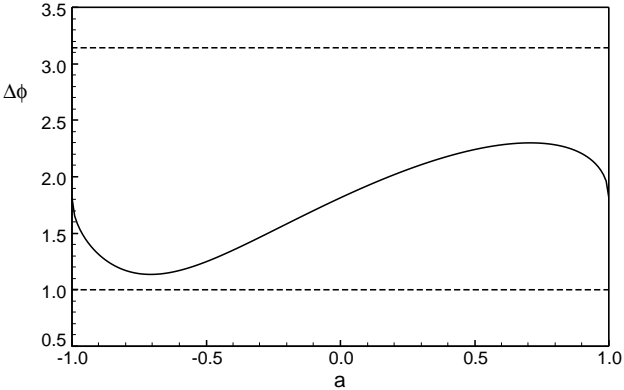}
\end{center}
\caption{$\Delta \phi$ vs. $a$}
\label{Fig:Dphi}
\end{figure}

Fig.~\ref{Fig:Dphi} shows that $\pi >\Delta \phi >1$ for all $-1\leq a\leq 1$
so that if the uncertainty relation holds for the root-mean-square deviation
$\Delta \phi $ then it also holds for $\left( \Delta \phi \right) _{S}=\pi $
as argued above.

\begin{figure}[H]
\begin{center}
\bigskip\bigskip\bigskip \includegraphics[width=9cm]{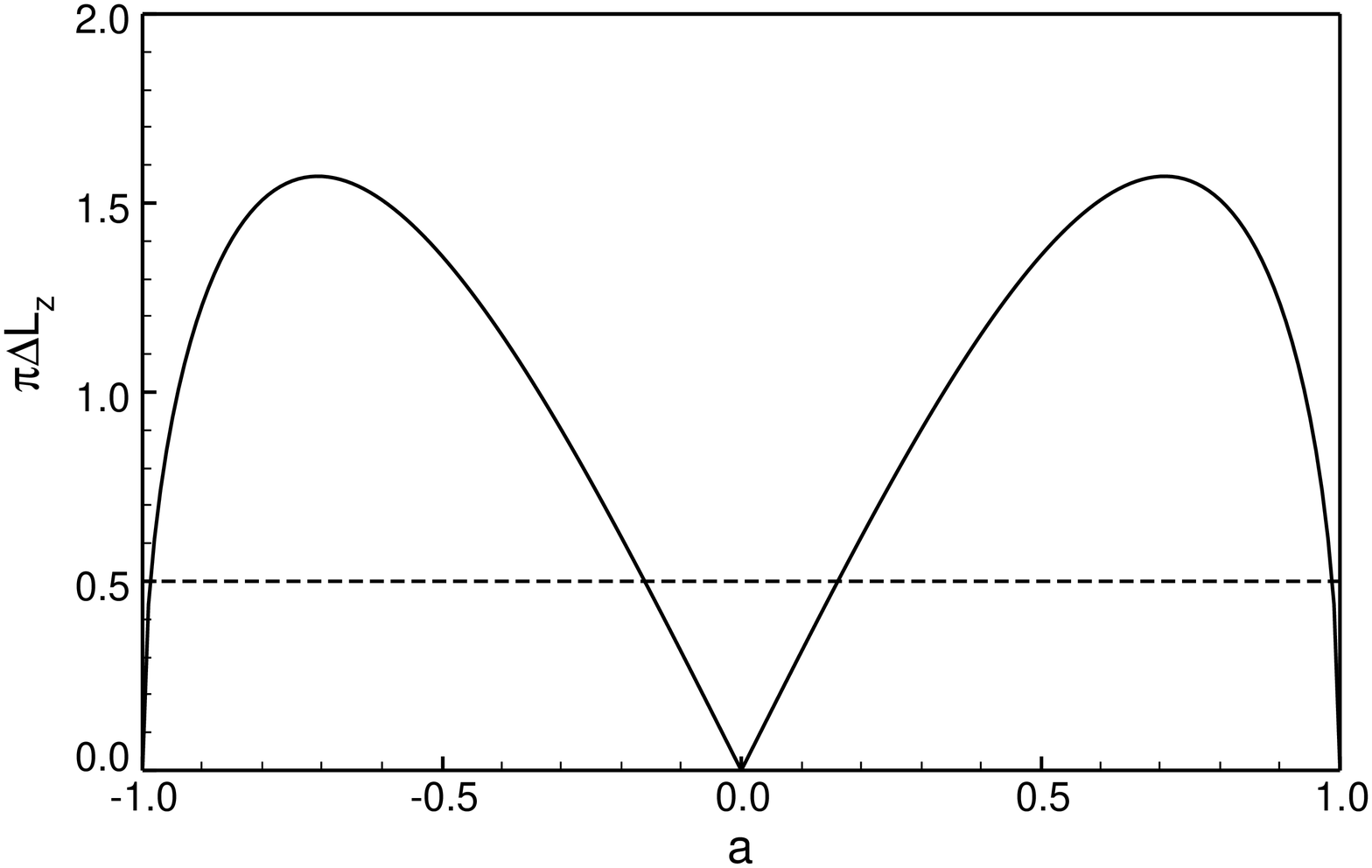}
\end{center}
\caption{$\pi \Delta L_z/\hbar$ vs. $a$}
\label{Fig:piDLz}
\end{figure}

Fig.~\ref{Fig:piDLz} shows that $\pi \Delta L_{z}=\hbar /2$ at
four points $a_{1}^{\prime }=-a_{4}^{\prime }\approx -0.99$ and
$a_{2}^{\prime }=-a_{3}^{\prime }\approx -0.16$ and that $\pi
\Delta L_{z}\geq \hbar /2$ for $a_{1}^{\prime }\leq a\leq
a_{2}^{\prime }$ and $a_{3}^{\prime }\leq a\leq a_{4}^{\prime }$.
This uncertainty relation fails for $a$ outside those intervals.
We appreciate that the inequality invoked by Strange\cite {S12}
(which he arbitrarily chose to be $\Delta \phi \Delta L_{z}\geq
\hbar $) is not valid for all possible states of the system.

For simplicity we have restricted the discussion of the
uncertainty relation to states that depend only on the azimuthal
angle. By no means does such restriction invalidate the
conclusions drawn from the state (\ref {eq:Psi(a,phi)}) that are
illustrated in figures \ref{Fig:DphiDLz}, \ref {Fig:Dphi} and
\ref{Fig:piDLz}. However, as a further pedagogical exercise it is
worth taking into account the actual motion of the electron on the
$x-y $ plane. If we repeat the calculation for states $f(r,\phi
)=\left\langle r,\phi \right| \left. f\right\rangle $ and the
inner product
\begin{equation}
\left\langle f\right| \left. g\right\rangle =\int_{0}^{\infty
}\int_{0}^{2\pi }f(r,\phi )^{*}g(r,\phi )r\,d\phi \,dr
\end{equation}
we obtain the exact uncertainty relation
\begin{equation}
\Delta \phi \Delta L_{z}\geq \frac{\hbar }{2}\left| 2\pi \rho (2\pi
)-1\right|  \label{eq:Delta_phi_Delta_Lz_r_phi}
\end{equation}
where
\begin{equation}
\rho (\phi )=\int_{0}^{\infty }\left| \psi (r,\phi \right| ^{2}r\,dr
\label{eq:rho(phi)}
\end{equation}
Equation (\ref{eq:Delta_phi_Delta_Lz_r_phi}) is a generalization
of the uncertainty relation (\ref{eq:Delta_phi_Delta_Lz}) that was
derived earlier by Kraus\cite{K65,K70}. Note that equation
(\ref{eq:Delta_phi_Delta_Lz_r_phi}) is suitable for the $(r,\phi
)$-dependent states chosen by Strange\cite {S12} to illustrate the
probability backflow. For example, using Strange's three-term
wavefunction (his equation (11) properly normalized)\cite{S12} we
obtain $\Delta \phi \Delta L_{z}\approx 1.99\hbar $ and
$\frac{\hbar }{2}\left| 2\pi \rho (2\pi )-1\right| \approx
0.844\hbar $ that satisfy the uncertainty relation
(\ref{eq:Delta_phi_Delta_Lz_r_phi}). Exactly in the same way we
can easily generalize the uncertainty relations derived by
Chisolm\cite{C01} that provide tighter lower bounds to the
products of square-root-mean deviations.

\section{Conclusions}

\label{sec:conclusions}

In this paper we have shown that the uncertainty relation invoked
by Strange\cite{S12} in his discussion of the probability backflow
is only valid for a particular set of wave functions. The electron
in a constant magnetic field is a suitable example for showing
that the $\phi -L_{z}$ uncertainty relation should be applied
carefully because it is different from the $x-p$ one. In order to
keep the discussion as simple as possible we have avoided more
complicated issues like the correct form of the operator for the
azimuthal angle and of its square-root-mean
deviation\cite{J64,K65,PT69,K70}. Instead, we have kept the most
straightforward definitions of both the operator $\hat{\phi}$ and
its square-root-mean deviation $\Delta \phi $\cite {C01} that
proved suitable for the analysis of the results obtained by
Strange\cite{S12}.

Finally, we point out that in the case of the motion of a particle
in three dimensions one can easily derive uncertainty relations
similar to equation (\ref{eq:Delta_phi_Delta_Lz_r_phi}) that
generalize those derived earlier by other
authors\cite{K65,K70,C01}.

\begin{thebibliography}{9}
\bibitem{S12}  Strange P 2012 \textit{Eur. J. Phys.} \textbf{33} 1147.

\bibitem{J64}  Judge D 1964 \textit{Nuovo Cim.} \textbf{31} 332.

\bibitem{K65}  Kraus K 1965 \textit{Z. Physik} \textbf{188} 374.

\bibitem{PT69}  Perlman H S and Troup G J 1969 \textit{Am. J. Phys.} \textbf{%
37} 1060.

\bibitem{K70}  Kraus K 1970 \textit{Am. J. Phys.} \textbf{38} 1489.

\bibitem{P79}  Peslak Jr. J 1979 \textit{Am. J. Phys.} \textbf{47} 39.

\bibitem{C01}  Chisolm E D 2001 \textit{Am. J. Phys.} \textbf{69} 368.
\end{thebibliography}
\end{document}